\documentclass[prl,twocolumn,groupedaddress,nofootinbib]{revtex4}
\usepackage[bookmarks=true,bookmarksopen=true,pdfhighlight=/I,pdfpagemode=UseOutlines]{hyperref}
\usepackage[latin1]{inputenc}
\usepackage{graphicx}
\usepackage{amssymb,amsmath}
\usepackage[usenames,dvipsnames]{color}
\usepackage{hyperref}

\usepackage{amsmath}
\usepackage{amssymb}
\newcommand{\nin}{\noindent}
\newcommand{\be}{\begin{equation}}
\newcommand{\ee}{\end{equation}}
\newcommand{\bea}{\begin{eqnarray}}
\newcommand{\eea}{\end{eqnarray}}
\newcommand{\br}{\hskip .25cm/\hskip -.25cm}

\newcommand{\nn}{\nonumber\\}

\newcommand{\ovl}{\overline}

\begin{document}

\nin CERN-PH-TH-2011-073, KCL-PH-TH/2011-{\bf8}

\title{Mass Hierarchies in Lorentz-Violation-induced Dynamical Mass Models}

\author{Jean Alexandre}
\affiliation{King's College London, Department of Physics, Strand, London WC2R 2LS, UK.}

\author{Nick E. Mavromatos}
\affiliation{CERN, Theory Division, CH-1211 Geneva 23, Switzerland.\\ 
King's College London, Department of Physics, Strand, London WC2R 2LS, UK}

\begin{abstract}
We discuss the issue of fermion mass hierarchies between lepton and quark families in toy models of dynamical generation of fermion masses through minimal Lorentz Invariance Violation (LIV). Realistic hierarchies in chiral theories necessitate an extended gauge structure, which includes both vector and axial-vector interactions.
\end{abstract}

\maketitle

\emph{\textbf{Introduction.}} In \cite{vergou} a model with minimal violation of Lorentz invariance (LIV), associated with a mass scale $M$, has been put forward
as a way of dynamically generating fermion masses. In \cite{mavromatos}, such models have been embedded in
more microscopic quantum gravity theories, such as multi-brane world models with space-time defects, in which cases the scale $M$ is close to the Planck scale,
for instance, in the models of \cite{mavromatos}, $M \sim M_s/g_s$, where $M_s$ is a string scale, and $g_s $ a
perturbatively weak string coupling, $g_s \ll 1$. Such embeddings enhance the dynamically generated masses, which in a simple minded field theoretic setting appear unrealistically small, to phenomenologically realistic values via an inverse Randall-Sundrum hierarchy fashion~\cite{RS}. In \cite{mavromatos} a discussion on the gauge invariance of the physical dynamical mass
and how the latter can be obtained through appropriate resummation techniques in field theory, such as the pinch technique~\cite{pinch}, has also been presented. In the models of \cite{vergou,mavromatos} only one type of fermions (representing electrons or, more generally, a charged lepton) have been considered. It is the point of this brief note to extend the model to incorporate different kinds of fermions, e.g. charged leptons and quarks, and discuss ways of generating mass hierarchies among them. We shall consider massless gauge fields. The delicate issue of generating masses for gauge bosons through Lorentz violation, as an alternative to Higgs mechanism, will be postponed to a future publication.

\emph{\textbf{Review of LIV Vector Interactions.}} We commence our analysis by first reviewing  some properties of the LIV QED model of \cite{vergou}, consisting of a vector Abelian
$U_V(1)$ interaction. The lagrangian is given by
\be\label{model1}
-\frac{1}{4}F_{\mu\nu}\left( 1-\frac{\Delta}{M^2}\right) F^{\mu\nu}+\ovl\psi \left( i\br\partial-g_V\br A \right) \psi~,
\ee
where $g_V < 1$ the (perturbatively weak) coupling of the vector interactions, $M$ is a scale indicating the
strength of the Lorentz Invariance Violation (LIV) and $\Delta = \vec\partial \cdot \vec\partial$ is the spatial
Laplace operator.
The lagrangian (\ref{model1}) is invariant under the vector $U_V(1)$ gauge group:
\be\label{vectorgt}
\psi \rightarrow e^{i g_V \theta } \psi ~, \quad A_\mu \rightarrow A_\mu + \partial_\mu \theta~,
\ee
and the vector propagator is
\bea\label{D}
&&D_{\mu\nu}^{bare}(\omega,\vec p)= \nonumber \\ && -\frac{i}{1+p^2/M^2}\left( \frac{\eta_{\mu\nu}}{-\omega^2+p^2} +
(\xi -1)\frac{p_\mu p_\nu}{(-\omega^2+p^2)^2}\right)~, \nn
\eea
where $p^0= \omega$ and $p^2=\vec p\cdot\vec p$. We thus observe that, because the pole structure is not
affected  by the Lorentz Violating (LV) terms and thus the photon remains massless in this minimally LV model.

The one-loop effective action calculated in \cite{vergou} in the fermion sector reads (in the conventions of that reference):
\bea\label{oneloopfermi}
&& S_{{\rm eff~fermi}} = \nonumber \\ & &\int d^4 x ~\overline{\psi} \left( i[1 + Z_0] \gamma^0 D_0
- i[ 1 + Z_1] \vec{\gamma} \cdot \vec{D} -m_{\rm dyn}\right) \psi ~, \nonumber \\
& &  D_\mu = \partial_\mu + i g_V A_\mu ~,
\eea
where $m_{\rm dyn}$ is the fermion dynamical mass, as explained in the next subsection, and
\begin{eqnarray}
Z_0 & = & - \frac{\alpha}{2\pi} \left( {\rm ln}\left(1/\mu\right) + 4 {\rm ln}2 - 2 \right) + \mathcal{O}\left(\mu^2 {\rm ln}(1/\mu)\right)~, \nonumber \\
Z_1 & = & - \frac{\alpha}{2\pi} \left( {\rm ln}\left(1/\mu\right) + \frac{50}{9} - \frac{20}{3}{\rm ln}2 \right) + \mathcal{O}\left(\mu^2 {\rm ln}(1/\mu)\right)~, \nonumber \\
\mu & \equiv & \frac{m_{\rm dyn}}{M}~.
\end{eqnarray}
In the limit $m_{\rm dyn} \ll M$, which characterizes theories in which the Lorentz violation is due to
the quantum gravity sector,
as we mentioned previously, with $M$ close to the Planck scale, we observe that the dominant terms in the
wave function renormalization functions $Z_0$, $Z_1$ are equal, and thus the fermion sector assumes an
approximately relativistic form.
As explained in \cite{vergou}, after rescaling time and $A_0$ one can show that the gauge sector remains gauge invariant
and that the speed of light is 1. Indeed, the fermion dynamical mass being different from 0, the speed of light cannot be
defined from the fermion dispersion relation. The speed of light is defined from the gauge field dispersion relation,
which keeps it stadard relativistic
form. Also, the derivation of the Ward identity is independent of the details of the photon bare propagator, and has
its standard relativistic form
\be\label{WI}
k_\mu\Gamma^\mu(p,k)= g_V G^{-1} (p) - g_V G^{-1}(p-k)~,
\ee
where $\Gamma^\mu$ and $G$ depend on $p^0$ and $\vec p$, but not through the relativistic combination $p_0^2-(\vec p)^2$.
In a dynamically generated scenario with only attractive boson interactions, using the above-mentioned fact that
the fermion/gauge sector maintains (approximately) its Lorentz-invariant form,
one writes
\be\label{fermi}
G^{-1} (p) = \mathcal{A}(p) \gamma^\mu p_\mu + \Sigma(p) + \cdots
\ee
where dots denote LIV terms~\footnote{In general, one could have written:
$$ \mathcal{A}_0 (p^0, \vec{p}) \gamma^0 p_0
+ \sum_{i=1}^{3} \mathcal{A}_i (p^0, \vec{p}) \gamma^i p_i $$ which however,
to one loop and to leading order in $\mu$ may be well approximated by (\ref{fermi}).}.
This form of Ward identity was used in \cite{mavromatos} in order to argue
on the applicability of pinch techniques~\cite{pinch} to guarantee that the gauge invariant contribution
of the (lowest order) dynamically generated fermion mass in \cite{vergou} was the one corresponding
to the Feynman gauge $\xi =1$, in which only the longitudinal parts of the gauge boson propagator
enter the physical scattering matrix amplitudes.

The vector interaction is \emph{attractive}, and thus, following the standard trend on superconductivity-like
scenarios for dynamical formation of fermion condensates and opening of energy gaps, one can have dynamical
formation of fermion masses.
The Schwinger-Dyson (SD) equation for the fermion propagator read:
\be\label{SD}
G^{-1}-G_{bare}^{-1}=g_V\int D_{\mu\nu}\gamma^\mu G\Gamma^\nu,
\ee
where $\Gamma^\nu$, $G$ and $D_{\mu\nu}$ are respectively the dressed vertex, the dressed fermion propagator
and the dressed photon propagator.
To lowest order, $\Gamma^\nu = g_V\gamma^\nu $, while the LIV vector $U_V(1)$ boson propagator is given by eq.(\ref{D}).

Neglecting the loop corrections to the photon propagator, as well as the fermion wave function renormalization,
that is keeping only the corrections to the electron self-energy, the dressed fermion propagator can then be expressed as:
\be\label{G}
G(\omega,\vec p)=i \,\frac{ p_\mu \gamma^\mu + m_{\rm dyn}}{p_\mu p^\mu - m_{\rm dyn}^2 },
\ee
where $m_{\rm dyn}$ is the fermion dynamical mass.

With these approximations, the SD equation (\ref{SD}) involves a convergent integral,
due to the $M$ dependent Lorentz-violating terms.
In \cite{mavromatos} has been argued, using pinch technique arguments, that the GFP independent value corresponds to the
Feynman gauge, where the gap equation is
\bea\label{gapeq}
&&m_{\rm dyn} =  \\ 
&&\frac{\alpha_V}{\pi^2}\int_{-\infty}^\infty  d\omega \int_0^\infty \frac{p^2dp}{1+\frac{p^2}{M^2}}
\frac{4m_{\rm dyn}}{(\omega^2+p^2)(\omega^2+p^2+m^2_{{\rm dyn}})}~, \nonumber
\eea
with $\alpha_V=g_V^2/4\pi$. This equation has the trivial solution $m_{\rm dyn}=0$, which is not preferable energetically,
and the non-vanishing solution
\be\label{mdyn}
m_{\rm dyn} \simeq M \exp\left(-\frac{\pi}{2\alpha_V}\right)~.
\ee
For weak couplings $\alpha_V \ll 1$ this is much smaller than $M$, and the solution is self consistent.

To leading order, one may ignore the wave function renormalization, so the dynamical mass corresponds to the value
$\Sigma (p^2=m_{\rm dyn})$ in eq.(\ref{fermi}). In fact, from the Ward identity (\ref{WI}) and (\ref{fermi}) it follows
that in the limit
$k \to 0$, the right-hand-side vanishes, even in the presence of a non-zero scalar $\Sigma (p)$.  This implies that
in the vector
$U_V(1)$ gauge theory there are no poles of the dressed vertex function as $k \to 0$. This is consistent with the
absence of spontaneous breaking of the $U_V(1)$ gauge theory.
The above argument that there is no dynamical mass generation for Abelian vector bosons
parallels the Lorentz Invariant (LI) case~\cite{pagels}.

\emph{\textbf{Chiral Vector Gauge Bosons.}} Let us now extend this model to include chiral vector fields, that is gauge/fermion vertices involving the $\gamma_5$ Lorentz matrix.
The presence of a $\gamma_5 \gamma^\mu $ fermion/gauge boson vertex introduces anomalies, which can be canceled if and only if one assumes a collection of fermionic fields $\psi = (\psi_1,\cdots,\psi_n)$
coupled to an axial-vector (A) field $B_\mu$ as follows:
\begin{equation}\label{chiral}
\mathcal{L}_A = -\frac{1}{4} G_{\mu\nu} \left( 1-\frac{\Delta}{M^2}\right) G^{\mu\nu}+\ovl\psi \left( i\br\partial-g_A \br B \gamma_5 \underline{\tau} \right) \psi~,
\end{equation}
with $g_A < 1$ a weak axial-vector coupling, $G_{\mu\nu} = \partial_\mu B_\nu - \partial_\nu B_\mu$ and $\underline{\tau}$ is an $n \times n$ hermitean matrix, satisfying the following condition that ensures the absence of the triangle anomaly AAA and thus the renormalizability of the model
\be\label{cond}
{\rm tr}\{\underline\tau\} = 0
\ee
In what follows we shall restrict ourselves to fermion doublets for brevity. We shall use
\be\label{tau3}
\psi = \begin{pmatrix} \psi_1 \\  \psi_2 \end{pmatrix}~, \qquad \underline{\tau} \equiv \underline{\tau}_3
= \begin{pmatrix} 1 & 0 \\ 0 & -1 \end{pmatrix}
\ee
although other choices can be made~\cite{massivevector,massivevector2}, as long as the anomaly-free condition (\ref{cond}) is satisfied.

The action (\ref{chiral}) has an axial vector  $U_A(1)$ symmetry:
\be\label{axialgt}
\psi \rightarrow e^{i g_A \gamma_5 \theta} \psi ~, \quad B_\mu \rightarrow B_\mu + \partial_\mu \theta~.
\ee
The purely axial vector interactions are \emph{repulsive}, as follows by the relevant Schwinger-Dyson equation for the fermion self-energy to lowest order (\emph{c.f}. (\ref{SD}))
when one replaces the vector $\Gamma^\mu $ with a dressed chiral vertex function $\Gamma^\nu_5 = \gamma^\nu \gamma_5 + \dots $ to lowest order. As a result of the anti-commutativity $\{ \gamma_5 , \gamma^\mu \} = 0$, one obtains
a solution of the form (\ref{mdyn}), but with $\alpha_V$
replaced by $-\alpha_A=-g_A^2/4\pi$. Thus, for weak couplings $\alpha_A \to 0$ this would diverge
which is inconsistent.
This is in agreement with the repulsive nature of the axial vector interactions.

\emph{\textbf{Vector and Axial-Vector $U_V(1) \otimes U_A(1)$ Theories.}} We next consider a model that combines two gauge symmetries $U_V(1) \otimes U_A(1) $.
The model has a fermion  singlet $\chi$ which couples with a coupling $\tilde g_V$ only
to a vector symmetry, with gauge bosons $A_\mu$, and a fermion doublet $\psi = (\psi_1,\psi_2) $
couples to both vector $U_V(1)$ gauge bosons, $A_\mu$, with a coupling $g_V \ll 1$ and axial-vector $U_A(1)$
gauge bosons, $B_\mu$, with coupling $g_A \ll 1$.\\
The pertinent Lagrangian reads:
\bea\label{model}
&& -\frac{1}{4}F_{\mu\nu}\left( 1-\frac{\Delta}{M^2}\right) F^{\mu\nu}
-\frac{1}{4} G_{\mu\nu} \left( 1-\frac{\Delta}{M^2}\right) G^{\mu\nu}\\
&&+ \ovl\chi \left( i\br\partial-{\tilde g}_V\br A \right) \chi +
\ovl\psi \left( i\br\partial-g_V\br A -g_A \br B \gamma_5  \underline{\tau}_3\right) \psi ~, \nonumber
\eea
where the $\tau_3$ matrix is the $2 \times 2$ Pauli matrix given in (\ref{tau3}) and this structure is necessary
for $\gamma_5$ AAA anomaly cancelation.
This model is not anomalous if $g_V \ne 0$, in the sense that anomalous graphs containing odd powers of $\gamma_5$
cancel out upon the condition (\ref{cond}). The inclusion of non-Abelian groups yields in general anomalies for chiral
fermions, but one hopes that an overall anomaly cancelation, and thus renormalizability, will be restored
in phenomenologically realistic models involving non-Abelian standard-model-like groups, to which $U_V(1) \otimes U_A(1)$
may be extra factors (see discussion below).
The model (\ref{model}) has the following gauge invariances:
\bea
U_V(1) &:& A_\mu \rightarrow A_\mu + \partial_\mu \theta~, \quad \psi \rightarrow e^{i g_V \theta} \psi, \quad
\chi \rightarrow e^{i {\tilde g}_V \theta} \chi \nonumber \\
U_A(1) &:& B_\mu \rightarrow B_\mu + \partial_\mu \varphi ~,\quad
\psi \rightarrow e^{i g_A \gamma_5 \varphi} \psi ~ .
\eea
For the moment, one is tempted to assume that in the model, the fermion $\chi$ to represent a charged lepton,
with $\tilde g_V = e$, thus the unbroken vector $U_V(1)$ gauge symmetry representing electromagntism,  while
the doublet $\psi$ represent a quark family, with
\be\label{gtildeg}
g_V = r {\tilde g}_V ~, \quad r < 1~,
\ee
representing the fractional charge of quarks (all these refer to bare values, assuming that these equal the
value of the corresponding running couplings at the appropriate renormalization points). Only the quarks are
assumed to couple to the chiral $U_A(1)$. The issue is to see weather we are able to generate
an electron/quark hierarchy dynamically. In a realistic model, whose construction is not our aim here, the
lepton $\chi$ would have been paired with its neutrino. However, since the neutrino is electrically neutral,
its coupling to the electromagnetic $U_V(1)$ would have been suppressed by higher loop effects of the standard
model. This would generate a natural hierarchy of mass scales between neutrinos and the rest of the leptons
\cite{mavromatos}. We ignore such complications in our toy model studied in this article.

\emph{\underline{Fermion self energy structure}. } The fermions $\psi$ are exposed to competing forces, attractive ones due to the vector gauge boson interactions, with structure constant $\alpha_V = g_V^2/4\pi$, and repulsive axial vector interactions, with fine structure constant $\alpha_A = g_A^2/4\pi $.
Naively, if we followed the analysis in \cite{vergou}, we could then write down the analogue of the SD equation (\ref{SD})
for the fermion $\psi$ self energy, with the fermion propagator
\be
G^{-1}(p) = \br p + \Sigma (p)
\ee
to leading order.
The issue here is the flavour structure of the gap function $\Sigma (p)$, which can generate a mass for the vector
boson $B_\mu$ as we now explain. \\
We first note that the axial Ward identity maintains its relativistic form (its derivation is independent of the details
of the bare vector propagator):
\be\label{aWI}
k_\mu\Gamma_5^\mu(p,k)=g_AG^{-1}(p)\gamma_5 \underline\tau_3-g_A\gamma_5 \underline\tau_3 G^{-1}(p-k)~,
\ee
where $\Gamma_5^\mu$ and $G$ depend on $p^0$ and $\vec p$, but not through the relativistic combination $p_0^2-(\vec p)^2$.
Then we decompose the fermion self energy in the form
\be\label{Sigma}
\Sigma(p) = \Sigma_s (p) + \underline\tau_2 \gamma_5 \Sigma_v (p)~,
\ee
where both $\Sigma_s$ and $\Sigma_v$ are scalar in fermion flavour space and the matrix $\underline\tau_2$ satisfy
$ [\underline\tau_2, \underline\tau_3 ] = 2i\underline\tau_1~.$

From eqs.(\ref{aWI}) and (\ref{Sigma}), one can see that if $\Sigma_v\ne0$,
the dressed axial vertex $\Gamma_5^\mu$ has a pole for vanishing vector momentum $k^\mu$, since
\be\label{wiaxialf}
\lim_{k\to0}~ k_\mu\Gamma_5^\mu(p,k) \ne0~,
\ee
As explained in \cite{massivevector,massivevector2}, this pole is responsible for the dynamical generation of a mass for the vector
boson $B_\mu$. It is not clear, however, that the system chooses a finite flavour-changing self energy
contribution $\Sigma_v$, as we now explain. \\
In refs.~\cite{massivevector,massivevector2}, where a non-vanishing bare fermion mass has been assumed, it was demonstrated that
the Schwinger-Dyson equation for the fermion propagator is consistent with the following asymptotic form for $\Sigma_v$
\be\label{asymptot}
\lim_{p^2\to\infty}\Sigma_v(p)\sim \kappa (p^2)^{-\epsilon}~,~~~~\mbox{with}~~0<\epsilon<1~,
\ee
where $\kappa$ is a dimensionful {\it arbitrary} parameter. As a consequence, the dressed model consists of a
{\it continuous} set of theories. Since physical quantities should be independent of the continuous parameter
$\kappa$, it can be expected~\cite{massivevector2} that the vacuum energy is \emph{the same} for the cases of massive and massless bosons $B_\mu$. However, the asymptotic behavior (\ref{asymptot}) is introduced in order to get a finite gap equation, whose solution depends on $\epsilon$. The limit $\epsilon \to 0$ is then taken in \cite{massivevector2} in the so called ``platform'' approximation,
according to which $\epsilon/(\alpha_V-\alpha_A)$ remains finite. In this way the vacuum energy for the massive $B_\mu$ boson case
\emph{may} be lower than the one for the massless case, and thus one \emph{may} obtain a dynamical $B_\mu$-mass generation~\cite{massivevector,massivevector2}. But this was not demonstrated rigorously so far, as it involves non-perturbative considerations.

In our situation, however, dynamical $B_\mu$-boson mass generation \emph{may not} happen for the following reason:
in our approach we do not use a regularization of the fermion self energy in the form of a power law
$\kappa (p^2)^{-\epsilon}$, since the gap equation we obtain is finite, because of the mass scale $M$.
In this sense, the arbitrary continuous parameter $\kappa$ of ref.~\cite{massivevector2} does not arise, and therefore it is not clear weather the argument given in that work holds. As a consequence, for our purposes here we assume that the vacuum energy for the massless $B_\mu$ boson case is lower than the corresponding energy in the massive $B_\mu$ case, and thus dynamical $B_\mu$-mass generation does not take place. This issue is, however, a delicate one, and we postpone its detailed analysis in a forthcoming publication.
Nevertheless, even if there is an energetically preferred ``Higgs-less'' broken phase for the vector field $B_\mu$, this will not affect qualitatively the fermion hierarchies discussed below, although undoubtedly it will complicate technically the solution of the associated Schwinger-Dyson equations.

\emph{\underline{Fermion dynamical masses and hierarchy}.} According to our previous discussion, the fermions $\chi$ can 
have dynamical mass generation, to lowest order in a perturbative expansion for the coupling ${\tilde g}_V \ll 1$ which 
assumes the form
\be\label{mchi}
m_\chi \simeq M \exp\left(-\frac{\pi}{2{\tilde \alpha}_V}\right) ~~,~\mbox{with}~~
{\tilde \alpha}_V = {\tilde g}_V^2/4\pi~.
\ee
The vector $U_V(1)$ remains unbroken for reasons discussed above and this implies the massless-ness of the vector boson
$A_\mu$ in the model and this is why we identify it with ordinary electromagnetism in this toy model.

As far as the fermion $\psi$ is concerned, it can easily be seen that the relevant gap equation is
\bea\label{dm2}
&&m_\psi = \frac{\alpha_V - \alpha_A}{\pi^2}\times\\
&&\int_{-\infty}^\infty  d\omega \int_0^\infty \frac{p^2dp}{1+ \frac{p^2}{M^2}}
\frac{4m_\psi}{(\omega^2+p^2)(\omega^2+p^2+m^2_\psi)}~,\nonumber
\eea
which, if $\alpha_V > \alpha_A$ (the vector attraction dominates over the axial vector repulsion),
has the physically consistent solution
\be\label{mpsi}
m_\psi \simeq M \exp\left( -\frac{\pi}{2({\alpha}_V-\alpha_A)}\right) ~,  \quad \alpha_V > \alpha_A~.
\ee
A scalar mass of the fermions breaks the axial gauge symmetry of the action (\ref{model}).

Comparing eqs.(\ref{mchi}) and (\ref{mpsi}),
we observe that it is possible to generate a mass hierarchy between
the fermion families $\chi$ and $\psi$:
\be\label{hierarchy}
\frac{m_\chi}{m_\psi}
= \exp\left(\frac{\pi}{2} \times\frac{(1 - r^2)\alpha_V + r^2 \alpha_A}{\alpha_V \left(\alpha_V - \alpha_A\right)}\right)~.
\ee
For $\alpha_V$ sufficiently close to $\alpha_A$ we may generate large mass hierarchies,
but unfortunately in the wrong way if the vector coupling of the lepton $\chi$ is larger than
that of the quark family ($r < 1$).\\
On the other hand, if we do not view the vector $U_V(1)$ as the electromagnetic one, but consider
the $U_V(1) \otimes U_A(1)$ gauge group as an extra (to the Standard Model group) Abelian factor
group, \emph{e.g}. of the type generated in some string theories~\cite{leontaris}, then we may take $r > 1$ and
still the real electric charge of the quarks and leptons given by the Standard Model group,
which is assumed not to have LIV terms.
In such a case, a phenomenologically correct mass hierarchy is obtained if
$(1 - r^2) \alpha_V + \alpha_A < 0$, in which case the ratio (\ref{hierarchy}) is smaller than 1.
The absolute masses, although appearing unrealistically small in such models,
nevertheless can be enhanced by embedding the model in Randal-Sundrum-type~\cite{RS} multibrane
scenarios, as discussed in \cite{mavromatos}.

\section*{Acknowledgements}

N.E.M. acknowledges discussions with G. Leontaris, and J.A acknowledges discussions with K. Farakos and A. Tsapalis. 
This work is partially supported by the Royal Society (UK).


\begin{thebibliography}{99}

\bibitem{vergou} J.~Alexandre,
  [arXiv:1009.5834 [hep-ph]];
J.~Alexandre and A.~Vergou,
  arXiv:1103.2701 [hep-th].

\bibitem{mavromatos} N.~E.~Mavromatos,
  Phys.\ Rev.\  {\bf D83}, 025018 (2011).

\bibitem{RS} L.~Randall and R.~Sundrum,
  Phys.\ Rev.\ Lett.\  {\bf 83}, 3370 (1999).


\bibitem{pinch} J.~M.~Cornwall,
  Phys.\ Rev.\  D {\bf 26}, 1453 (1982);
J.~M.~Cornwall and J.~Papavassiliou,
  Phys.\ Rev.\  D {\bf 40}, 3474 (1989);
For a review see: D.~Binosi and J.~Papavassiliou,
  Phys.\ Rept.\  {\bf 479}, 1 (2009).



\bibitem{pagels} R.~Stern,
  Phys.\ Rev.\  D {\bf 14} (1976) 2081;
see also: H.~Pagels,
  Phys.\ Rev.\  D {\bf 21}, 2336 (1980).


\bibitem{massivevector}
R.~Jackiw and K.~Johnson,
  Phys.\ Rev.\  D {\bf 8} (1973) 2386;

\bibitem{massivevector2} J.~M.~Cornwall and R.~E.~Norton,
  Phys.\ Rev.\  D {\bf 8} (1973) 3338.

\bibitem{leontaris} See, for instance:  G.~K.~Leontaris and J.~Rizos,
  Nucl.\ Phys.\  B {\bf 567}, 32 (2000).




\end{thebibliography}
\end{document}